\documentstyle[prl,aps,twocolumn]{revtex}
\begin{document}
\sloppy
\draft
\title{Low-temperature conduction and DC current noise in a
quantum wire with impurity}
\author{Ulrich Weiss   }
\address{Institut f\"ur Theoretische Physik, Universit\"at Stuttgart,
D-70550 Stuttgart, Germany     }
\date{\today}
\maketitle
\begin{abstract}
The nonlinear conductance for tunneling through an impurity in a
Luttinger liquid and the nonequilibrium DC current noise are 
calculated exactly for low temperatures. 
We present a pedestrian pathway towards the exact solution 
which is based on analytic properties and on a duality between 
weak and strong backscattering. The prefactor of the
$T^2$ enhancement is shown to be universally expressed in terms of
zero temperature properties. 
\end{abstract}
\pacs{PACS numbers: 72.10.-d, 73.40.Gk    }
\narrowtext
Many-body correlations are essential in 1D electron 
systems, where the usual Fermi liquid behavior is destroyed by the 
interaction. The generic features of many  1D interacting fermion
systems are well described in terms of the Luttinger liquid model
\cite{luther,emery}.  
In the Luttinger model, all effects of the electron-electron interaction
are captured by a single dimensionless parameter $g$ which is in
the range $g<1$ for repulsive interaction. A sensitive experimental
probe of a Luttinger liquid state is the tunneling conductance
through a point contact in a 1D quantum wire \cite{kafi92}. 
Of interest is also the DC nonequilibrium current noise \cite{fls95b}.
Tunneling of edge currents 
in the fractional quantum Hall (FQH) regime provides another
realization of a Luttinger phase. As shown by Wen \cite{wen}, the edge
state excitations are described by a (chiral) Luttinger liquid with
$g=\nu$, where $\nu$ is the fractional filling factor.

In this paper, we make use of analytic properties and 
of a duality between weak and strong backscattering, which allow
us  to recover the exact solution of the Luttinger liquid tunneling
\cite{fls95a} and the DC current noise problem \cite{fls95b}
at $T=0$. We then study low $T$ and compute the prefactor of the 
$T^2$ enhancement exactly.

The low-energy modes of the 1D interacting electron liquid are 
conveniently treated in the frame work of standard bosonization 
\cite{luther,emery}. 
The creation operator for spinless fermions can equivalently
be expressed in terms of boson phase fields $\theta(x)$ and $\phi(x)$,
which obey the equal-time commutation relation
$[\phi(x),\theta(x')]= -(i/2) {\rm sgn}(x-x')$
[we put $\hbar=k_B=e=1$]. 
The clean Luttinger liquid is described by the Hamiltonian
\[
H_L = \frac{v_F}{2 g}\int dx\,[(\partial_x\theta)^2/g + 
g(\partial_x\phi)^2 ]\; .
\] 
We assume a sharp band width cutoff
$\omega_c$ for the linear dispersion relation implicit in $H_L$. 
 
Backscattering with $2 k_F$ at a short-ranged impurity potential 
of strength $V_0$ is described by the generic form
\[
H_I = - V_0 \cos[2\sqrt{\pi}\theta(0)] + V \theta(0)/\sqrt{\pi} \; .
\] 
We have added a term describing an applied voltage drop $V$
at the impurity. 
The $\theta$ representation is convenient for the discussion of 
weak backscattering. 

For the opposite limit of a large barrier,
the $\phi$ representation is more appropriate. Disregarding multiple
electron hops, the barrier and voltage contribution takes the form
($\Delta$ is the overlap matrix element)
\[
H_I' = -\Delta \cos[2\sqrt{\pi}\phi(0) +Vt] \; .
\]
Here, $\phi(0)$ is the discontinuity of the $\phi$ field at the 
barrier. The harmonic liquid is again described by $H_L$.

In the following, we study the nonlinear static conductance
$G(V) = I(V)/V$ through an impurity described by $H_I$ or $H_I'$.  
Using the bosonized form of the current operator in
the $\theta$ representation, we have
$G_\theta(V) = \lim_{t\to\infty} \langle \dot{\theta}(0,t)\rangle
/\sqrt{\pi} V$, where $\langle\cdots\rangle$ denotes the thermal
average over all modes of $H_L$ away from $x=0$.

The model $H_L +H_I$ is equivalent to the problem of a 
Brownian particle of mass $m$ moving in a cosine potential
$V_{WB}(q) = -V_0\cos(2\pi q/q_0)$
\cite{schmid,guinea,w88,w93}, referred to as the
WB model.
The correspondence can be shown by canonical 
transformations and examination of the equations of motion of
the coordinate and momentum 
autocorrelation functions. The quantity $\theta(0)$ corresponds to 
$\sqrt{\pi} q/q_0$.
Ohmic damping is provided by excitation of the
harmonic liquid away from the barrier, and the parameter $g$ is related
to the Ohmic viscosity $\eta$ by $1/g=\alpha\equiv\eta q_0^2/2\pi$. 
The cutoff frequency $\omega_c$ of the liquid modes is identical
with $\eta/m$. The equivalence becomes exact when the force of
inertia is negligibly small compared with the friction force. For 
$T\ll V_0$ and $V\ll V_0$, this means 
$\eta^2 \gg m V_{WB}''(0)$, 
or equivalently $\omega_c \gg 2\pi g V_0$.

By writing down a real-time path integral for the 
conductance or mobility \cite{w93,w91},
then formally expanding the exponent of the action in powers of 
$V_0^2$, and for each term integrating out $\theta(x,\tau)$, 
the resulting expression is the analog of a statistical 
ensemble of interacting discrete charges.  
Introducing the normalized conductance
$\widetilde{G}(V,T,g) =G(V,T,g)/G_0(g)$, 
where $G_0(g) = g/2\pi$ is the microwave conductance of the
1D quantum wire in the zero frequency limit \cite{kafi92},
we have for the $\theta$ model
\cite{schmid,w95b}
\begin{equation}\label{conduc1}
\widetilde{G}_\theta (V,T,g) = 1 - (\pi/V) {\rm Im}\,U(V,T,g)\; .
\end{equation}
The function $U$ describes the interacting charge gas,
\begin{eqnarray}\label{uvt}
 U(V,T,g) &=& \sum_{n=1}^\infty (iV_0)^{2n} \int_0^\infty
d\tau_1\,d\tau_2\cdots d\tau_{2n-1} \\ \nonumber
&& \times \sum_{\{\xi\}} W_n \prod_{j=1}^{2n-1}\left[ e_{}^{-
igVp_{j,n}\tau_j}
\sin(\pi g p_{j,n} )\right] \; , \\   \label{intfac}
W_n &=& \exp\bigg( \sum_{j>k=1}^{2n} \xi_j S(\tau_{jk})\xi_k \bigg) 
\; .
\end{eqnarray}   
The interaction $S(\tau)$ is the boson two-point function, 
\begin{equation}\label{stau}
 S(\tau) = 2g \ln [(\omega_c/\pi T)\sinh(\pi T \tau)] \; .
\end{equation}
The $2n-1$ integration times $\tau_j$ in order $V_0^{2n}$ are the
intermediate
times between the $2n$ successive charges ($\xi_j=\pm
1;\;j=1,\cdots,2n$), 
and $\tau_{jk}$ is the distance between charges $\xi_j$ and $\xi_k$.
In every order $n$, the $2n$ charges have zero total charge,
$\sum_{j=1}^{2n} \xi_j =0 $. There are only arrangements of charges
for which the cumulative charge quantities 
$p_{j,n} = \sum_{i=j+1}^{2n} \xi_i$
are nonzero since otherwise the corresponding phase factor
$\sin(\pi g p_{j,n} )$ would vanish.

By proper identification of the parameters, (\ref{conduc1})
-- (\ref{stau}) is identical with the series expansion of
the normalized mobility of the WB model for vanishing inertia.

Similarly, the strong barrier $\phi$ model $H_L +H'_I$
is identical to a dissipative periodic tight-binding (TB) model, 
as shown by  mapping the Hamiltonians onto 
each other \cite{guinea}. To proceed, we recollect an exact duality 
transformation between observables of the dissipative WB and TB
model holding for Ohmic \cite{schmid} and for 
frequency-dependent damping \cite{w95a}.
Equally, there holds an exact duality between 
observables of the $\phi$ and $\theta$ model. Importantly, the
conductances $\widetilde{G}_\theta(V,T;V_0,g)$ and 
$\widetilde{G}_\phi(V,T;\Delta,g)$ 
of these models are related by 
\begin{equation}\label{duality}
\widetilde{G}_\theta(V,T;V_0, g) = 1 - 
\widetilde{G}_\phi(gV,T;V_0, 1/g) \; ,
\end{equation}
where we displayed the relevant parameters. 
For repulsive interaction, $g$ is restricted to 
$g < 1$. If we work on the related WB or TB model, the 
applicable interval is $0 < g < \infty$.

Putting $T=0$, the series (\ref{conduc1})
is a function of only $(V/T_0)^{2g-2}$, where $T_0 = a_\theta(g)
V_0^{1/(1-g)} \omega_c^{-g/(1-g)}$ is the characteristic scale 
introduced by the impurity. The prefactor $a_\theta(g)$ will be
specified shortly. In terms of $T_0$, the series may be 
written in the form
\begin{equation}\label{expan1}
\widetilde{G}(V,g) = 1 -\sum_{n=1}^\infty c_n(g)
\left(\frac{V}{T_0}\right)^{2(g-1)n}  \; .
\end{equation}
The coefficient $c_n(g)$ is given as a ($2n-1$)--fold integral.
The circle of convergence of (\ref{expan1}) is $v(g)<1$, where 
\begin{equation}\label{vg}
v(g) = (V/T_c)^{g-1}\quad \mbox{with} \quad  T_c = f(g) T_0 \; .
\end{equation}
The energy scale $T_c(g)$ is a crossover parameter analogous to the
Kondo temperature. It is determined below.  

There follows from (\ref{duality}) that the conductance is a power
series of $(V/T_0')^{2/g-2}$ in the complementary range $v(1/g)<1$. 
Demanding that the $\phi$ and $\theta$ model are different 
representations of the same physical problem, the scale $T_0'$ 
is identical with $T_0$. Thus we have 
\begin{equation}\label{expan2}
\widetilde{G}(V,g) = \sum_{n=1}^\infty c_n (1/g)
\left(\frac{V}{T_0}\right)^{2(1/g -1)n}  \; .
\end{equation}

Importantly, (\ref{expan1}) and (\ref{expan2}) are
different expansions of the same function with complementary regions
of convergence. They apply to the $\theta$ and
$\phi$ model. For $g<1$, (\ref{expan1}) represents
the high-voltage or weak-backscattering expansion, while (\ref{expan2}) is 
the expansion for low voltage or strong backscattering. 
Conversely, for $g>1$, the parameter regions for the series (\ref{expan1})
and (\ref{expan2}) are interchanged. As a result of strict duality, 
the entire expansions around weak and strong backscattering 
are related.

Kindly, the dual expansions (\ref{expan1}) and (\ref{expan2})
of the nonlinear conductance impose sufficiently strong 
conditions that it is possible to find the 
analytic form of the corresponding $T=0$ scaling function.

An appropriate integral representation which creates the dual series
expansions (\ref{expan1}) and (\ref{expan2}) reads 
\begin{equation}\label{intrep}
 \widetilde{G}(V,g) = \frac{1}{2\pi i} \int_{\cal C} \frac{dz}{z} 
\Gamma[u(z)] \Gamma[w(z)]F(z) 
\left(\frac{V}{T_0}\right)^{2z} \;,
\end{equation}
where $u(z)= 1 +z/(1-g)$, $ w(z)=1+zg/(g-1)$ and $F(0)=1$.
Now assume that $F(z)$ does not depend on $g$, and is an entire 
function of $z$. The integrand in (\ref{intrep}) is an
analytic function over the entire complex plane save for the points 
$z=z_n^{(L)} \equiv (g-1)n$ [$n=0,\,1,\,2,\ldots$] and 
$z=z_m^{(R)} \equiv (1/g - 1)m$ [$m=1,\,2,\ldots$] where it
possesses simple poles. For $g<1$, the contour ${\cal C}$ starts 
at $-i\infty$ and ends at $i\infty$, and circles 
the origin such that the set of poles $\{z_n^{(L)}\}$  
lie to the left, and the set of poles
$\{z_m^{(R)}\}$  lie to the right of the integration path. For
$g>1$, the path is in reverse direction and the origin is to the left.

For $v(g) < 1$, we close the contour such that all poles to the 
left are circled. Taking into account the residua of the enclosed 
poles, we find the series (\ref{expan1}) with
\begin{equation}\label{coeff1}
c_n(g) = (-1)^{n-1}\, g\Gamma(gn)F[(g-1)n]/\Gamma(n)\; .
\end{equation}

In the complementary range $v(1/g) < 1$, the contour ${\cal C}$ 
is closed by circling all poles to the right of the integration 
path. We then get the series expansion (\ref{expan2}) with 
(\ref{coeff1}).
 
Next, we utilize the assumption that $F(z)$ does not depend on $g$.
It is then possible to determine $F(z)$ by matching the expansion
(\ref{expan1}) or (\ref{expan2}) onto the known solution for 
$ g= 1/2$. The mobility in analytic form of the $g = 1/2$ case 
has been derived first in Refs. \cite{w88,w91} for any $V_0$ and $T$.
For $T=0$ and $g=1/2$, we get from (\ref{conduc1}) and
(\ref{uvt}) 
\begin{equation}\label{g12}
\widetilde{G}(V,1/2) = 1 - (T_0/2V) \arctan (2V/T_0)\; .
\end{equation}  

Upon matching the power series of (\ref{g12}) for $T_0 > 2V$ onto
the expansion (\ref{expan2}) with (\ref{coeff1}) at $g=1/2$, we find
\begin{equation}\label{fz}
F(z) = \Gamma(3/2) / \Gamma(3/2 + z) \; ,
\end{equation}
which indeed is analytic in the entire complex plane.
The same form is found if the series 
of (\ref{g12}) for $T_0 < 2V$ is matched onto (\ref{expan1}).
With (\ref{fz}), the coefficient (\ref{coeff1}) reads
\begin{equation}\label{coeff2}
c_n (g) = (-1)^{n-1} 
\frac{\Gamma(gn+1)\Gamma(3/2)}{\Gamma(n+1)\Gamma[3/2 + (g-1)n]} \; .
\end{equation}

Readily, the crossover scale $T_c (g) $ delimiting
the regions of convergence of the expansions (\ref{expan1}) and
(\ref{expan2}) follows from the requirement that the integrand in
(\ref{intrep}) tends to zero faster than $1/|z|$ on the semicircle
of the respective path of integration for $|z|\to \infty$.
With (\ref{coeff2}), we get from (\ref{intrep})
\[
T_c(g) = \sqrt{|1-g|}\, g_{}^{g/[2(1-g)]}T_0 \; . 
\] 
Interestingly, we have $T_c(g) =T_c(1/g)$. This follows from 
the invariance of the integrand in (\ref{intrep}) under 
$g \to 1/g$.

The $T=0$ expressions (\ref{expan1}) and (\ref{expan2}) with
(\ref{coeff2}) agree with the corresponding ones 
derived by Fendley {\it et al.} \cite{fls95a}. They utilized
a suitable basis of interacting quasiparticles in which the model is 
integrable, and they employed sophisticated thermodynamic Bethe-ansatz
(TBA) technology to calculate the non-Fermi distribution function
and the density of states  of the quasiparticles,
which then determine the conductance by a Boltzmann-type rate
expression. Our energy scale $T_0$ agrees with their scale $T_B'$. 
The much simpler derivation given here sheds additional light on the 
underlying symmetries of the models considered.

It is important to note that the representation 
(\ref{intrep}) with (\ref{fz}),
and the respective expansions (\ref{expan1}) and (\ref{expan2}) 
are exact for
$0<g<\infty$ under assumption that the correlation function $S(\tau)$ 
is described by the $T=0$ form of (\ref{stau}).

To relate the energy scale $T_0$ in the exact solution to the
parameters of the
$\theta$ model, we match the $n=1$ term of the series expansion 
(\ref{expan1}) onto the $n=1$ term of the series (\ref{conduc1}) with
(\ref{uvt}) for $T=0$. Thus we find
\begin{equation}\label{t1}
T_0^{2-2g} = g_{}^{2g-2} 2_{}^{2-2g} [\pi/\Gamma(g)]^2
V_0^2 \omega_c^{-2g}\; .
\end{equation}
Equation (\ref{intrep}) with (\ref{t1}) is an exact integral 
representation of the conductance of the $\theta$ model 
(and the equivalent WB model) at $T=0$ for any $V$ and 
any positive $g$.

Likewise, we obtain the conductance in the $\phi$ model by
expressing $T_0$ in terms of the parameters of this model. 
Upon matching the $n=1$ term of the series (\ref{expan2}) onto
the term of order $\Delta^2$ of the $\phi$ model, we find
\begin{equation}\label{t2}
T_0^{2-2/g} = 2_{}^{2-2/g} [\pi/\Gamma(1/g)]^2
\Delta^2 \omega_c^{-2/g} \; .
\end{equation}

Note that (\ref{t1}) and (\ref{t2}) are symmetric under the 
substitution $g\to 1/g$ except for the factor $g^{2g-2}$ in 
(\ref{t1}). It is just this extra factor which is the reason 
for the substitution rule $V \to V/g$ in the duality relation 
(\ref{duality}).

From (\ref{t1}) and (\ref{t2}) we find the functional
dependence between $V_0$ and $\Delta$,
\[
\Delta^2 = [\Gamma(1+1/g)]^2[\Gamma(1+g)]^{2/g}
(\omega_c/\pi)^{2+2/g} V_0^{-2/g} \; .
\]
Solving this for $V_0^2$ gives the same functional form except 
that $g$ is replaced by $1/g$. By use of this relation, 
we can equivalently express the $n=1$ term of (\ref{expan2}) 
for the $\phi$ model in terms of the corrugation strength $V_0$,
\begin{eqnarray}\nonumber
G_\phi^{(1)}(V,g) &=& [\pi/2\Gamma(2/g)]\,\Delta^2 V^{2/g -2}
\omega_c^{-2/g} \; ,\\  \label{instanton}
 &=& \frac{g^{2/g-2}\Gamma(1/g)[\Gamma(g)]^{2/g}}{\sqrt{\pi}\Gamma(1/2+1/g)}
\,\frac{\omega_c^2 V^{2/g-2}}{(2\pi V_0)^{2/g}} \; .
\end{eqnarray}
While the first form is of order $\Delta^2$ and is the simple
golden rule expression of the conductance in the $\phi$ model, 
the second form is the exact single-bounce or instanton-pair 
contribution in the $\theta$ model, which is nonperturbative in
$V_0^2$. Until now, the single-bounce contribution in the $\theta$
model has been calculated only for $1/g=\alpha\gg 1$ 
\cite{korshunov}. The expression
(\ref{instanton}) agrees with the result of Ref.\cite{korshunov} in this 
limit. The expression (\ref{instanton}) is the generalization
for all $g$, but $\omega_c \gg 2\pi g V_0$.
Interestingly, the single-bounce contribution as well as all
multi-bounce terms in the series (\ref{expan2}) have been found without 
calculating bounce actions and tricky fluctuation determinants. 

For $g=1/2 - \varepsilon$, with $|\varepsilon|\ll 1$, the expressions
(\ref{expan1}) and (\ref{expan2}) with (\ref{t1}) agree with the
findings in Ref.\cite{w95b}, where a leading-log summation of the 
corrections about the $g=1/2$ solution was made.

For $g=1 - \varepsilon$ with $|\varepsilon|\ll 1$, the series
(\ref{expan1}) or (\ref{expan2}) gives
\begin{equation}\label{c1}
\widetilde{G}_\theta (V, 1-\varepsilon) = 
 1/[1 + (\pi V_0/\omega_c)^2(2\omega_c/V)^{2\varepsilon}]  \; .
\end{equation}
This form agrees with the result by Matveev, Yue, and Glazman
\cite{matveev}, who performed a leading-log summation for 
weak electron-electron interaction.

The limit $\alpha = 1/g \to \infty$ represents the classical limit in 
the related Brownian particle model. For $g=0$, the series
(\ref{expan1}) is summed to the form
\begin{equation}\label{c2}
\widetilde{G}_\theta (V,0) = 
\sqrt{1 - (2\pi V_0/V)^2} \; ,\quad  V\ge 2\pi V_0\; .
\end{equation}    
This expression is indeed the solution of the associated Fokker-Planck
equation in the Smoluchowski limit \cite{risken} for $T=0$. In the
region $V< 2\pi V_0$, the conductance is zero because the slope of 
the tilted washboard potential $H_I$ changes regularly sign, and the 
overdamped classical particle comes to rest at a point with zero slope.

The conductance as a function of $V/T_0$ shows a smooth transition from
the square root singular behavior (\ref{c2}) via the kink-like shape
(\ref{g12}) to the constant behavior (\ref{c1}) as $g$ is increased from
$0$ to $1$. All curves cross the line $\widetilde{G} =1/2$ in the interval
$V/T_0= 2/\sqrt{3} \pm 0.01$.

Consider next the leading low-temperature correction. Taking into
account the $T^2$ contribution of the correlation function (\ref{stau}),
the interaction factor (\ref{intfac}) reads 
\begin{equation}\label{wtemp}
W_n = W_n^{(0)}\bigg[ 1 - g T^2\frac{\pi^2}{3} \bigg(\, \sum_{j=1}^{2n-1}
p_{j,n}\tau_j \bigg)^2\, \bigg] \; ,
\end{equation}
where $W_n^{(0)}$ is the full interaction factor at $T=0$. 
There follows from (\ref{uvt}) that the annoying factor 
$(\sum_j p_{j,n}\tau_j)^2$ in (\ref{wtemp}) may be generated by 
differentiation with respect to the voltage \cite{w91}. Thus, the 
prefactor of the $T^2$ correction is universally expressed in terms 
of the nonlinear conductance at $T=0$. For $T\ll gV$, we have
\begin{equation}\label{tempcorr}
G(V,g;T) = G(V,g) + \frac{\pi^2 T^2}{3gV} \,\frac{\partial^2}{
\partial V^2} \,[VG(V,g)] \;.
\end{equation}
Unfortunately, $T^4$ and higher order corrections cannot
be expressed in such simple terms.

Finally, consider the nonequilibrium DC current noise $\delta I^2(V,T)$.
Within the interacting quasiparticle picture, exact expressions
for the DC noise at any $V$ and $T$ involving
TBA integral equations have been given and evaluated numerically
in Ref.\cite{fesa96}. Within the Coulomb gas representation, 
it is straightforward to write
down the series expressions of the noise at any $V$ and $T$ \cite{w91}. 
For the $\theta$ model, comparison of the series of the expression 
\begin{displaymath}
\delta I^2_\theta(V,T) \equiv \lim_{\omega\to 0} \int dt\, e^{i\omega t}
\langle \{\dot{\theta}(0,t),\dot{\theta}(0,0)\}\rangle /2\pi  \;.
\end{displaymath} 
with the series (\ref{conduc1}) yields at $T=0$
\begin{equation}\label{noise1}
\delta I_\theta^2 (V) = - (g V/2) V_0 (\partial/\partial V_0) 
G_\theta (V)   \; .
\end{equation}
For a weak barrier, only the term of order $V_0^2$ 
in (\ref{uvt}) is relevant. We then have 
$\delta I^2_\theta(V) = - g V G_\theta^{(1)}(V) $.
In the FQH device, this noise is due to quasiparticle tunneling.

Similarly, one finds for the $\phi$ model, either by direct calculation
or by use of (\ref{noise1}) and the duality relation (\ref{duality}),
\begin{equation}\label{noise2}
\delta I_\phi^2 (V)   = (V/2) \Delta (\partial/\partial\Delta)
 G_\phi (V) \; .
\end{equation}
For a strong barrier, the leading term is of order $\Delta^2$, yielding
$\delta I^2_\phi(V) =    V G_\phi^{(1)}(V) $. In the FQHE effect,
this noise results from tunneling of uncorrelated electrons.

Since $G_\theta$ is a function of $V_0^2 V^{2g-2}$, and $G_\phi$ is a 
function of $\Delta^2 V^{2/g -2}$, we may express both (\ref{noise1})
and 
(\ref{noise2}) as
 \begin{equation}\label{noise3}
\delta I^2(V) = [g/2(1-g)] V^2 (\partial/\partial V) G(V)
\end{equation}
These results for the noise at $T=0$ agree with those of Ref.
\cite{fls95b}, where it was however derived quite differently.

Knowing the DC noise at $T=0$ exactly, allows us to 
calculate the exact $T^2$ enhancement as well. Taking up the 
arguments based on (\ref{wtemp}), we obtain for $T \ll gV$ and in 
the regime $v(g) <1$, in which (\ref{expan1}) is appropriate,
\begin{displaymath}
 \delta I^2(V,T) = \delta I^2(V) 
+ \frac{\pi^2T^2}{3g}\, \frac{\partial^2}{\partial V^2} \delta
I^2 (V) + 2G_0 T \; .
\end{displaymath}
Clearly, this form applies both for the $\theta$ and the $\phi$ model. 
Importantly, the prefactor of the $T^2$ enhancement of the current 
noise is again described by zero temperature properties. 
The additional term $2G_0T$ is the familiar Johnson-Nyquist noise 
which corresponds to the Einstein relation between diffusion 
coefficient and mobility in the related Brownian particle 
model \cite{w93}. In the complementary regime $v(1/g)<1$,
we have the same expression for $\delta I^2(V,T)$ except that
the last term is absent, because the series (\ref{expan2}) applies.
 
To conclude, we have computed the conductance and DC current noise
in a Luttinger liquid with a source of backscattering at low $T$.
A pedestrian pathway founded on duality and analytic properties
guided us to the analytic solutions at $T=0$. 
Strict duality means that the entire expansions around weak and
strong backscattering are related. In the FQHE device, the crossover
from weak to strong backscattering comes with a crossover from
Laughlin quasiparticle tunneling  to electron tunneling. 
We calculated in analytic form the leading 
low-temperature enhancement for arbitrary strength of the 
impurity potential.

I wish to thank M. Sassetti for previous collaborations and
interesting conversations.

\end{document}